


\input harvmac


\Title {\vbox {\baselineskip14pt{\hbox{CERN-TH.6820/93}}
\hbox{hep-th/9303054}  }}
{\vbox{\centerline {Two--dimensional Conformal  Sigma Models } \vskip2pt
\centerline{
 and  Exact  String   Solutions}
}}
\centerline{ A.A. Tseytlin\foot{
On leave from P.N. Lebedev Physics Institute, Moscow, Russia and
Blackett Laboratory, Imperial College, London SW7 2BZ, U.K.} }
\centerline {\it Theory Division, CERN}
\centerline {\it
CH-1211 Geneva 23, Switzerland}

\bigskip

\centerline {\bf Abstract}
\medskip
\noindent
\baselineskip 8pt
We discuss  two classes of   exact (in $\alpha'$) string
solutions  described by conformal sigma models. They can be viewed as  two
possibilities of constructing a conformal model out of the non-conformal
one based on the metric of a $D$-dimensional homogeneous $G/H$ space.
The first   possibility  is  to introduce two extra dimensions
(one space-like and one time-like) and to impose the null Killing
symmetry condition on the resulting $2+D$ dimensional metric.
In the case when the ``transverse" model is $n=2$ supersymmetric
and the $G/H$ space is K\"ahler-Einstein  the resulting  metric-dilaton
background can be found explicitly.
The second  possibility -- which  is realised in the sigma models
corresponding to $G/H$ conformal theories -- is to deform the metric,
introducing at the same time a non-trivial dilaton and antisymmetric
tensor backgrounds. The expressions  for the  metric and dilaton in
this case are  derived using the operator approach in which
one identifies the equations for  marginal  operators of conformal
theory with the linearised (near a background) expressions for  the
sigma model  `$\beta$-functions'. Equivalent results   are then   reproduced
by the direct field-theoretical approach based  on computing first
the effective action of the  $G/H$ gauged WZNW model and then solving
for the $2d$ gauge field. Both the bosonic and  the supersymmetric
cases are discussed.
\bigskip
\noindent
 To be published in the  $\ $   {\it    Proceedings of
 INFN Eloisatron Project 26th  Workshop: ``From Superstrings to Supergravity",
}
 Erice - Sicily, 5-12 December 1992 (World Scientific, 1993)
\bigskip

\noindent
{CERN-TH.6820/93}

\Date
 {March 1993}

\vfill\eject

\def \G {\Gamma} \def \k1 {{1\over
k}} \def \bh { {\bar h} } \def \ov { \over }

\def \O {\Omega }
 \def \del {\partial }
\def \ra {\rightarrow}
\def \W { {\bar \o }}
 \def \in {\int d^2 z  }
\def \half{{\textstyle{1\over 2}}}

\def \D {\Delta}

\def \a {\alpha}
\def \b {\beta}

\def \Tr {{\ \rm Tr \ }}

\def \ln {{\rm \ ln \  }}
\def \det {{\ \rm det \ }}

\def \1p {{1\over  \pi }}
\def \2p {{{1\over  2\pi }}}
\def \4p {{ {1\over 4 \pi }}}
\def \8p {{{1\over 8 \pi }}}
\def \P^* { P^{\dag } }
\def \p {\phi}

\def \m {\mu }
\def \n {\nu}
\def \ep {\epsilon}
\def\g {\gamma}

\def \k {\kappa }

\def \o {\omega}

\def \t {\theta}

\def \fourth {{\textstyle{1\over 4}}}
\def \third {{\textstyle{1\over 3}}}

\def \e#1 {{{\rm e}^{#1}}}
\def \const {{\rm const }}

\def \eq#1 {\eqno {(#1)}}
\def\inv{^{\raise.15ex\hbox{${\scriptscriptstyle -}$}\kern-.05em 1}}

\def\Dsl{\,\raise.15ex\hbox{/}\mkern-13.5mu D} 
\def\dsl{\raise.15ex\hbox{/}\kern-.57em\partial}
\def\del{\partial}

\def \sm {sigma model\ }

\def \dm {\del_\m}
\def \dn {\del_\n}
\def \go {G^{(0)}}

\def \bd  {{ \bar \del }}

\def \D  {\Delta }

\def \V { {\tilde V } }
\def \E {{ \tilde E}}
\def \bH {{\bar H}}

\def \dm {\del_\m}
\def \dn {\del_\n}
\def \go {G^{(0)}}

\def \bd  { \bar \del }
\def \D  {\Delta }

\def \ov {\over }
\def \kg { k + \ha c_G}
\def \kh { k + \ha c_H}
\def \A  { {\bar A} }
\def \tth {\tilde h}
\def \SWZNW {supersymmetric WZNW theory \ }
\def \C {{\cal C }}

\def \o {\omega}

\def \p {\phi}
\def \ep {\epsilon}

\def \m {\mu}
\def \g {\gamma}
\def \n {\nu}

\def \fourth {{1\over 4}}
\def \third {{1\over 3}}
\def \e#1 {{{\rm e}^{#1}}}
\def \const {{\rm const }}

\def \ggij {{g_{ij}}}

\def \hg {{\hat g}}

\def \tg {{\tilde g}}
\def \half { { 1\over 2 }}
\def  \ee {{\rm e}^}
\def \J {\bar J }
\def\np {  Nucl. Phys. }
\def \pl { Phys. Lett. }
\def \mpl { Mod. Phys. Lett. }

\def \pr  { Phys. Rev. }
\def \ap  { Ann. Phys. }
\def \cmp { Commun. Math. Phys. }
\def \ijmp { Int. J. Mod. Phys. }


 \vglue 0.8cm
\line{\bf 1. Introduction  \hfil}
\bigskip
As is well known,
solutions of  string field equations   are usually represented in
terms of conformal invariant $2d$ theories$^1$.
 Interpreting string theory as a theory of quantum gravity
one needs to study  solutions which have a  curved  space-time part, i.e.
which should be described by  conformal theories with Minkowski signature
of a target space.  It is important to  be able to go beyond the leading orders
of expansion
in $\a'$ (the leading order solutions may not   correctly describe
the behaviour in the strong curvature regions near
singularities, etc.).
Since the string effective  equations  contain terms of
arbitrarily high order in $\a'$ their  solutions  should    in general be
non-trivial
functions of $\a'$.
Given that   the exact  form of the string effective action is not known,
it may seem unlikely  that  such solutions can be found directly in the
perturbative sigma model framework.  It is   often  assumed   that  in contrast
to conformal field
theory methods,  the sigma model approach is useful  only  for obtaining
perturbative solutions   in
first leading  orders in $\a'$.

 Let us call `conformal sigma model' (c.s.m.) a $2d$  field theory with
couplings
that  solve the `$\b$-function' conditions of \sm conformal invariance,
reserving the name
`conformal field theory' (c.f.t.)  for a theory constructed using the operator
algebra (conformal
bootstrap) methods  (in  which case one usually knows  not only the  conformal
point but also the
spectrum of operators or perturbations).
The correspondence between c.s.m.'s and c.f.t.'s is  far
from being one-to-one and is poorly understood  in general.  Taking any
solution of the one-loop \sm
conformal invariance conditions, one can in principle  `deform'  it  order by
order in $\a'$ to get a
c.s.m.  for which in most cases the  corresponding c.f.t. will be unknown or
may not even exist
(e.g. if the $\a'$-series  does not converge, etc.).
There are also  explicitly   known  exact (all order in $\a'$) solutions  of
the  conformal
invariance conditions (in particular, with Minkowski signature) for which  the
existence of their
c.f.t.  counterparts  is unclear.  On the other hand, there are    many
c.f.t.'s which
apparently do not correspond to any weak coupling c.s.m., i.e. do not have an
obvious space-time
interpretation. Ideally, one would like to have both the  c.s.m. and c.f.t.
descriptions of  a
string solution to have an obvious space-time interpretation as well as to know
the spectrum of
states and to be able to compute their scattering  amplitudes on a given
background.

  We
would like  to describe here some recent  work  on    establishing    exact
string solutions directly in
$2d$ field theory or sigma model  framework.  In Sec.2 we shall briefly review
a  class of solutions
with Minkowski signature which generalise  an arbitrary non-conformal \sm in
$N$ dimensions to a
conformal one in $D=N+2$ dimensions  by replacing the `running' couplings  of
$N$-dimensional
theory by functions of  a light-cone coordinate in $N+2$ dimensions$^{2}$.  In
particular,
the explicit form of these exact solutions can be   found  in the
supersymmetric case$^3$.
Which conformal theories correspond to these solutions remains an open
question.

In the rest of the  paper we shall consider  a large  and important class of
(Euclidean or
Minkowski)  string  solutions  for which  the
 conformal field theory  description is well
known:   we shall  present a general  construction of  conformal  sigma models
corresponding to
coset $G/H$ conformal theories. Since coset  conformal field theories$^4$ can
be  represented at
the Lagrangian level  in terms of gauged WZNW models$^5$ one can try to obtain
a  \sm description
by  starting with  a gauged WZNW model, integrating out the $2d$ gauge field
and  fixing a gauge as
was first discussed for the $SL(2,R)/U(1)$ model in ref.6.  This gives,
however, only  the leading
order form of the \sm action since the quantum ($1/k$ or $\a'$) corrections are
  implicitly
ignored. In fact, one does find $\a'$-corrections to the  leading order
background$^{7,6}$ by
explicitly solving the  `$\b$-function' equations$^8$.  The exact form$^9$ of
the background metric
and dilaton can be  inferred indirectly  by using   the `operator approach'
(i.e. by comparing
the structure of the
 $L_0$-operator of  the  corresponding coset conformal theory with that of a
Klein-Gordon equation in a background) and turns out to be consistent with the
$\a'$-perturbation
theory for the sigma model$^{8,10}$.  This $SL(2,R)/U(1)$ solution$^9$
represents the first
explicitly known
 example of an exact string   solution which non-trivially depends on
$\a'$.\footnote{*}{\ninerm\baselineskip=11pt  As was  recently  observed$^{11}$
(see also
refs.12,13)  the causal structure of the exact solution is very different from
that of the leading
order  one$^6$, in particular, it does not have the `black hole' singularity.
The  exact solution in
the supersymmetric case, however,  still has the leading order `black hole'
form$^{10,13}$. \hfil}
The exact form of  some more general ($D>2$)  non-trivial $G/H$ backgrounds
was found in the
operator approach in ref.13.  We shall   describe  a direct field-theoretical
approach to derivation of  such    solutions in the general $G/H$
case$^{14,15,16}$.

The main idea of this  approach$^{14}$ is to first  find    the quantum
effective action of the
gauged WZNW  theory  and  then    eliminate the $2d$ gauge field.   After   the
resulting non-local
action is identified with the  effective action   of the corresponding  sigma
model, the  \sm
couplings  can be determined by dropping out all the non-local terms$^{16}$.
 Our presentation   will  follow mainly ref.16 (especially in Secs.4 and 5),
where
more details and references can be found.  In Sec.3 we shall  discuss,   from
a rather general
point of view, the operator approach to derivation of the exact  background
fields corresponding to
a conformal theory based on  an  affine-Virasoro construction.  In   Sec.4 we
shall present the
 expressions for the effective action in  the gauged WZNW theory (in  both
bosonic and
supersymmetric  cases).  The derivation of the \sm  couplings corresponding to
the $G/H$ gauged WZNW
theory  will be  given in  Sec.5.  In  Sec.6 we  shall establish the
equivalence   between   the
results  obtained in  the  operator  and   the \sm    approaches  and  make
some concluding
remarks.

\vglue 0.6cm
\line{\bf 2.   Conformal Sigma Models with Null Killing Vector \hfil}
\vglue 0.4cm
\def \ggij {{g_{ij}}}
\def \ha {{1\over 2}}
\def \hg {{\hat g }}
\def \gb {\beta}
\def \ga {\alpha}
Consider a non-conformal \sm in $N$ dimensions with  a metric $\ggij$.
One possible  strategy  of  constructing a conformal model out of it  could be
to couple it to a
$2d$  gravity with the conformal factor  of the $2d$ metric   becoming an
extra  $N+1$
coordinate$^{17}$.  The resulting conformal invariance equations  are, however,
 second order
in the new coordinate  and their solution is ambiguous (an `initial value'
problem is not well
defined).
 A much  simpler and  better defined  construction of a (Minkowski) conformal
\sm from a (Euclidean)
non-conformal one$^{2,3}$ is based on adding {\it two} extra coordinates  (say,
 light-cone
coordinates $u$ and $v$),   at the same time imposing the condition  of
$v$-independence. The
latter condition   (more precisely,  the  condition of existence of  a
covariantly constant  null
Killing vector)   can be thought  of as effectively reducing the dimension  to
$N+1$.  The conformal
invariance conditions in $2+N$ dimensions  then take the form of the standard
first order RG
equations
in $N$-dimensional theory with $u$ playing the role of the RG time.

 The general $D=N+2$ dimensional Minkowski signature metric
 admitting a covariantly constant null Killing
vector can
be represented in the form  $$ds^2 = \hg_{\mu \nu} dx^{\mu} dx^{\nu} =  -2dudv
+  \ggij (u,x) dx^i dx^j   \ \  ,
 \ \eq{1} $$ where $\mu $ and $  \nu = 0,1, ...,  N+1 , \
  i,j = 1,...,N . $
To establish  the UV finiteness of the  corresponding sigma model on a flat
$2d$ background one
should check that  there exists  a vector $M_\mu$  such that the $\gb$-function
 for the  target
space metric $G_{\mu \nu }=\hg_{\mu \nu}$ (1)  vanishes up to a
$M_\mu$-reparametrisation term
     $$ \gb^G_{\mu \nu} + 2 D_{( \mu} M_{\nu )} =0 \ . \eqno (2) $$
It  can be proved  that   (2)  is indeed satisfied for such   a $\ggij(x,u)$
as a function of $u$ that
$$ {d\ggij \over d u  } = \gb^g_{ij}  \ , \
  \eqno (3) $$
where $\gb^g_{ij}$ is the  $\gb$-function of the `transverse' ($N$-dimensional)
theory.
Namely,
the following  statement is true:  if the  metric $\ggij$
depends on $u$ in such a way that it satisfies the
standard RG equation of  the $N$-dimensional sigma model  then the
$(N+2)$-dimensional  sigma model
based on (1) is UV-finite to  all orders of the loop expansion.
This  statement  has a straightforward generalisation in the presence of
non-trivial dilaton, i.e.
for solutions of conformal invariance conditions on a curved $2d$ background:
 the metric  (1) and  the dilaton  $$\p = v +  \p (u,x) \ , \eq{4} $$
represent a solution of conformal invariance conditions in $N+2$ dimensions if
$\ggij (u,x)$ and $\p (u,x)$ solve the first-order RG equations of the
`transverse' theory. Given   a  set of couplings  $(\ggij (x) , \p (x))$
of a non-conformal theory in $N$ dimensions, one  obtains a conformal theory
in $N+2$ dimensions by solving the RG equations in $N$ dimensions and replacing
the RG parameter by the light cone coordinate $u$.

The metric  (and dilaton) in $N+2$ dimensions is   thus   determined   by the
$\gb$-function
of the transverse  theory.  The  explicit all-order expression for the latter
is not known in
bosonic sigma models.  On the other hand, there are examples of $n=2$
supersymmetric ($n$ is the number of $2d$ supersymmetries) sigma models  with
homogeneous  Einstein-K\"ahler target spaces  for which the  exact
$\gb$-function   coincides  with its    one-loop expression$^{18}$.
The above  general  statements   have direct
 analogues  in the case of supersymmetric sigma models$^3$.

For the metric with the null Killing  vector  (1)
 the  action of the  2-dimensional ($n=1$) supersymmetric  sigma
model can be represented  in
terms of the real superfields $U,\ V$ and $X^i$
$$ I= {1\over { 4 \pi  \ga'}} \int d^2 z d^2 \t \  [ -2 {\cal D} U { \bar {\cal
D} } V +
g_{ij}(U, X) {\cal D} X^{i} { \bar {\cal D} } X^{j} ] \  \ . \ \eqno
{(5)}$$
 The solution $g_{ij}(u,x)$ of the finiteness condition
is determined by the $\gb$-function of the `transverse'  part of
(5), i.e. of the supersymmetric model with the metric  $g_{ij}(u, x)$  for
constant $u$.  As is well known$^{19}$, if  the transverse space is K\"ahler
then the $N$-dimensional model is $n=2$ supersymmetric.  If
it is also a   compact  homogeneous Einstein space (e.g. $S^2= SO(3)/SO(2)$
or $CP^m$)  then   it is very plausible that its $\gb$-function is exactly
calculable  and is  given  by the one-loop expression$^{18}$. This   was
actually proved in ref.18 for the following classes of
 Einstein-K\"ahler manifolds:
  $
M_1= SO(m+2)/SO(m)\times SO(2)    $,  $
 M_2= SU(m+k)/SU(m)\times SU(k)\times U(1)  , $ $  M_3=Sp(m)/SU(m)\times U(1) ,
$  $   M_4=SO(2m)/SU(m)\times SO(2),   $ $ M_5 = SU(m+1)/ (U(1))^m .$  In that
case
the transverse part of the metric, the $\gb$-function    and the
solution of  the RG equation are  given simply  by
 $$ \ggij(u,x) = f(u) \gamma_{ij}(x)    \  , \ \ \ \ \gb (f) = a   \ , \ \ \
\  \  f(u) = a u  \ . \eqno (6) $$
The constant  $a > 0$  is  determined by the geometry of
the  transverse space$^{18}$ ($  a_1= m \  $, etc.)
 and  can be   absorbed  into  a redefinition
of the  coordinates $u$ and $v$. Then the final expression for the  Minkowski
signature
 metric of  the  finite  $2+N$-dimensional   supersymmetric
sigma model is (we  are  assuming  $u>0$)\footnote{*}{\ninerm\baselineskip=11pt
Note that while the transverse model (with fixed constant
$u$)  is $n=2$ supersymmetric,  the  full $(N+2)$-dimensional model
apparently  has  only $n=1$ supersymmetry.\hfil}
 $$ds^2  =  -2dudv
+  u \gamma_{ij }(x) dx^i dx^j   \ .  \ \eq{7} $$
 As a reflection of IR singularity of   the coupling of the transverse theory
this metric has
a curvature singularity at $u=0$.

The  simplest non-trivial example of the finite models we have constructed
corresponds to the case when the transverse theory is represented by the
$O(3)$ supersymmetric sigma model$^{20}$. The resulting metric
is that of $four$ ($=N+2$) dimensional space  with the transverse part
being  proportional to the metric  on $S^2$,
  $$ds^2  =  -2dudv
+  u (d\t^2 + {\rm sin}^2 \t d\varphi^2 )   \ \  .  \ \eq{8} $$
This  space  is conformal to
the  direct product of   two-dimensional Minkowski space and two-sphere.
To get a solution of the conformal invariance conditions one should
supplement the metric by the following
expression for the dilaton
$$\p (v,u)  =  \p_0 +  v + qu  +  {1 \over 8 } N
\ln u \ , \ \ \ \ \ q=\const \ . $$
The  resulting backgrounds   represent
exact solutions of superstring effective equations with {\it non-trivial}
dilaton$^3$.
 It is interesting to note  that the string coupling
$ e^{\phi} = A u^{N/8} e^{(qu +   pv)}\  $  goes to zero
 in the  strong coupling region $u \ra 0$
of the transverse sigma model, i.e. is {\it small} near the singularity $u=0$.
\vglue 0.6cm
\line{\bf 3.  Operator Approach to Derivation of Background Fields
 \hfil} \vglue 0.2cm
\line{\bf
 \  \ \ \ Corresponding to Coset Conformal Theories  \hfil} \vglue 0.4cm
A  possible strategy of
determining the  geometry  corresponding to a given conformal theory
is to try to
interpret the Virasoro condition $ (L_0 + {\bar L}_0 - 2) F  = 0$  on states as
 linear field
equations  in some background and
 to extract the expressions for background fields from the explicit form of the
differential operators involved.  The idea is that marginal operators $F$ of
conformal theory serve
as `probes' of geometry, so that one may be able to  extract  the corresponding
metric, etc from their
equations just as from  geodesic equations or field equations in a curved
space.
In order to implement this program one is  to make a number of important
assumptions.
First, one should  specify  which configuration (`target') space $M$  (with
coordinates $x^\m
, \ \m = 1,...,D$) should be used,  so that  $F$ will   be parametrised   by
fields on $M$, and $L_0$
acting on $F$  will reduce to differential operators  on $M$. Next, one  should
 understand how to
represent
 the resulting equations in terms of background fields.  The expectation is
that  the
conformal theory should correspond to a \sm
$$S = {1\over { 4 \pi \a' }} \int d^2 z \sqrt {\g} \ [      \del_m x^\m
\del^m x^\n G_{\m
\n}(x)    +  i \ep^{mn} \del_m x^\m \del_n x^\n  B_{\m\n}(x)    $$ $$  +  \a'
R  \p (x)
 +   T (x)  +  ...  ]  \  .
\eq{9} $$
One is to  invoke the knowledge of the
structure of the \sm   conformal  anomaly  coefficients (`$\b$-functions'), or
the effective action
which generates them
$$  S = \int d^{D}x \sqrt {\mathstrut G} e^{- 2
\phi} \lbrace {2 \over 3} (D-26) -  \alpha^{\prime} [R + 4
(\partial_{\mu} \phi)^{2}
- {1 \over 12} H^{2}_{\lambda \mu \nu}]
$$ $$  + {1 \over 16} [\alpha^{\prime} (\partial_{\mu} T)^{2} - 4T^{2}] +
\ldots
\rbrace \ .  \eq{10} $$
The idea is start with this background-independent action, linearise the
corresponding equations
near an arbitrary  background, and compare them  with the equations for the
corresponding states
in conformal theory.  The  equations for the tachyon,  graviton, dilaton   and
the
antisymmetric tensor perturbations
 ($t= T-T_*, \ h= G-G_* , \ \varphi = \p - \p_* , \ b = B- B_* $; in what
follows we shall omit
the superscript $*$ indicating background fields) take the following   symbolic
form  ($\a' = 1$)
$$  ( - \D + 2 G^{\m\n} \dm \p \dn)t  - 4t + ... =0     \ ,
\ \ \ \D \equiv  {1 \over  {\sqrt G} }
\del_\m (  {\sqrt G} G^{\m\n} \del_\n )\ ,\eq{11} $$
$$  ( - \D + 2 G^{\m\n} \dm \p \dn)h + Rh + H\del b  + ... = 0 \ , \eq{12} $$
$$ ( - \D + 2 G^{\m\n} \dm \p \dn)\varphi   +  H\del b  + R\del^2 h ... = 0 \ ,
 \eq{13} $$
$$  ( - \D + 2 G^{\m\n} \dm \p \dn)b +  H\del h + ... = 0 \ . \eq{14} $$
Given  a  second order differential equation  which follows from the
$L_0$-condition
for the lowest scalar `tachyonic' state
it  should  be
possible to determine the  corresponding  background metric and dilaton
 by looking at the coefficients of the terms which are second and first  order
in derivatives
   and comparing them with (11).   In an attempt to
determine the antisymmetric tensor field strength one  should   compare the
first-derivative terms
in the  equations for `massless' perturbations with  the corresponding terms in
(12),(13),(14).

It should be emphasised that if this approach works at all,
 its   consistency
should not be too surprising.  If the correspondence  between a  conformal \sm
and a conformal
field theory
exists  in a given case, then  solutions  of the conformal invariance
conditions that follow from
(10)   should represent the  conformal point; their perturbations   should
correspond to
marginal perturbations of a conformal theory.    {\it If}  there exists  a  \sm
(9) behind a given
conformal field theory, then the equations for marginal perturbations {\it
must} have  the form of
(11)--(13).

Let us now specialise to a large class of conformal theories based on
generalised  affine--Virasoro
construction$^{4,21,22,23}$.  Here one starts with
 a  finite-dimensional   Lie group $G$ and  defines a
holomorphic stress tensor
of a conformal theory  by  a Sugawara-type relation
$$ T_{zz}  = \C^{AB} :J_A(z) J_B(z):     \ \ , \ \eq{15} $$
where  $J_A (z) $  are
the generators of the  affine (current) algebra    determined by the structure
constants
$f^A_{\ BC}$  and the Killing metric $\eta_{AB}$ of $G$  (with the central term
proportional to
$\k_{AB}= k \eta_{AB}$). The condition that $T_{zz}$ should satisfy the
Virasoro
algebra imposes a
 `master equation'$^{21}$   on   the symmetric coefficients $\C^{AB}$
$$ \C^{AB} = 2 \C^{AC} \k_{CD}\C^{DB} - \C^{CD}\C^{KL} f_{CK}^A f_{DL}^B -
\C^{CD}f_{CL}^K f_{DK}^{(A} \C^{B)L} \  \eq{16} $$
(the central charge of the Virasoro algebra is $C= 2 \k_{AB} \C^{AB}$).
The standard Sugawara-GKO   $G/H$    coset conformal theory$^4$ corresponds to
a particular
solution of (16)
   $$ \C^{AB}=  {1 \over \kg } \eta^{AB}  - {1\over \kh}  \eta^{AB}_H
\ . \eq{17} $$
Here $ f^{ACD} f^B_{\ CD} = c_G \eta^{AB}\ , \ \   f^{acd} f^b_{\ cd} = c_H
\eta^{ab}$  ;   $\ A,B,...=1,..., {\rm dim}G=D_G$ ;
$a,b,...=1,..., {\rm dim} H=D_H$ and $\eta^{AB}_H $ denotes the projector on
the Lie algebra of $H$.

In general, unless solutions of (16)
 have non-trivial extra symmetries  (commuting operators),  the only
natural choice for a configuration space is the group space $G$  itself.
Representing the zero
modes of the currents  $J^A (z)$ and $\J^A (z)$  as differential operators  on
$G$
(with coordinates $x^M$)
$$  J_A=  E_A^M (x) \del_M \ , \ \ \ \J_A=  \E_A^M (x) \del_M \ , \eq{18} $$
$$ G_{0MN} =\eta_{AB} E^A_M E^B_{N}= \eta_{AB} \E^A_M \E^B_{N} \ , \ \ \  $$ $$
[J_A,J_B] = f^C_{AB} J_C \ , \ \ [\J_A,\J_B] = f^C_{BA}\J_C\ , \ \
[J_A,\J_B] = 0 \ ,  \eq{19} $$
 ($E_A^M$ and $E_A^M$ are the left-invariant  and right-invariant  vielbeins on
$G$;
 indices are raised and lowered with $\eta_{AB}$ and $ G_{0MN}$)
 we get   from  the zero mode part of $ (L_0 + {\bar L}_0 - 2) F  = 0$ the
following equation for the
lowest scalar
 state  $t(x)$
$$  [ - ( G^{MN} \del_M\del_N   +  G^N \del_N)  - 2] t (x) = 0 \ , \eq{20} $$
$$ G^{MN} =  \ha \C^{AB} (E^M_A E^N_B + \E^M_A \E^N_B) \ , \ \ \
  G^N  = \ha
    \C^{AB} (E^M_A \del_M E^N_B + \E^M_A \del_M \E^N_B) \ . \eq{21} $$
Eq.(20) becomes equivalent to  the \sm equation (11) if
there exists   a $\phi$ such that
$$ G^N = G^{MN} \del_M \ln (\sqrt G \ \e{-2 \p } ) \ . $$
In fact, such $\p$  can be  found explicitly by using  the properties of
$E^A_M$,
 or by observing that
that since   $J^A$ and $\J^A$   are   anti-Hermitian
with respect  to the  invariant scalar product on the group defined by $G_0$,
  $\ (f,g) = \int d^Dx {\sqrt{
G_{0}}} f^*(x) g(x) $,  one has
$$ \del_M E_A^M = - E_A^M \ \del_M  \ln {\sqrt {G_{0}}}\ , \ \ \ \
 E^A_N\del_M E^M_A = - E^{M}_A\del_M E^A_N\ ,  $$ $$ \ \ \
\del_M E^A_N - \del_N E^A_M = f^A_{BC} E^B_ME^C_N \ .  $$
As a result,
$$ \p =  \ha \ln  {\sqrt {G \ov G_{0}} }  \ , \  \ \
{\rm i.e.}
 \ \ \ \ \sqrt G \ \e{-2 \p} ={\sqrt {G_{0}}} \ . \eq{22} $$
This result becomes obvious if one compares the effective
action  leading to (11) ($ \int {\sqrt G} e^{-2 \p } G^{MN} \del_M t \del_N t
$)
with the `expectation value' $(t, Ht)$ of the zero-mode
`Hamiltonian' $H =\ha \C^{AB} (J_A J_B + \J_A \J_B) $
  and uses the antihermiticity of currents.
 The dilaton's role is to compensate for the fact that the two scalar products
have different measures.

 It is clear that the dilaton
field is  {\it  non-trivial}  because the metric $G_{MN}$ is,   in general,
different from the
canonical Killing metric $G_{0MN}$ on $G$.    A `deformation' of the metric is
directly related to
the conformal invariance (Virasoro) condition  (16).
If there exists the corresponding Lorentz-invariant \sm it should also contain
the antisymmetric
tensor field  coupling (cf. ref.24).

There seems to exist an  interesting  connection between algebraic   and
geometrical aspects of
such conformal theories  (and corresponding string   solutions).  The  geometry
is
determined by the choice of the group {\it and}  a  choice of particular
solution  of the
`master equation'.
 The  question about a relation between group-theoretic and
geometrical aspects  of a  similar   construction was raised independently  in
the
quantum mechanical
context in refs. 22  and  24. We see that  once the condition  of
conformal invariance
 is satisfied the geometry  that  appears  is
that of
the corresponding string solutions described by conformal sigma models.

Let us now specialise to the case of the $G/H$ coset conformal theory  with
$\C^{AB}$ given
by (17) where the main assumption of existence of the \sm description is
satisfied
given the existence of the Lagrangian formulation
 in terms of gauged WZNW models and since
this assumption
can be checked in the semiclassical approximation
(see also  refs.9,13).
 In this case there exists an extra  symmetry  which makes it possible
to   subject the states to the $H$-invariance condition $(J_a   - \J_a) F =0$.
In particular,
$$  (J_a  - \J_a) t  =0 \ , \ \ \   Z^M_a \del_M t = 0 \ , \ \ \ Z^M_a \equiv
E^M_a - \E^M_a
\ . \eq{23} $$
As a result, $t$ can be restricted to depend only on  $D=D_G - D_H$ coordinates
$x^\m$ of the coset
space $G/H$ which will thus play the role of the configuration space of the
corresponding sigma model.
The presence of the constraint (23) implies that  the metric we will get  from
(20),(21)
will be the `projected' one.
Let us   define
 the projection operator on the subspace orthogonal (with respect to $G_{0MN}$)
to $Z^M_a$
$$ \Pi^N_M
\equiv  \delta^N_M -  Z^N_a (ZZ)^{-1ab} Z_{Mb} \ , \ \ \ \ (Z
Z)_{ab} = G_{0MN} Z^M_a Z^N_{b}\ , \ \ \ \  \Pi^2 = \Pi \ . \eq{24} $$
Then
$$  G^{MN}=  \Pi^M_K {\hat G}^{KL} \Pi_L^N \ ,   \ \ \
{\hat G}^{MN}=  {1 \over \kg } \eta^{AB}E^M_AE^N_B   - {1\over \kh}  \eta^{ab}
E^M_aE^N_b \ ,
\eq{25} $$  $$ {\hat G}^{MN}={1
 \over \kg } ( E^M_AE^{AN} -  \g   E^M_aE^{aN}) = {1 \over \kg }[
E^M_iE^{iN}
-   (\g -1)   E^M_aE^{aN}] \ ,  \eq{26}   $$ $$
 \g =  {k+\ha c_G\over  k+\ha c_H } \ , \ \ \ \   \g - 1 ={ c_G - c_H\over 2
(k+\ha c_H) } \ .   \eq{27} $$
We have split  the  indices $A= (a,i) , \ i = 1,...,D$ on the indices
corresponding to the
subalgebra   and  the indices corresponding to the tangent space to $G/H$.  If
one  solves
(23) explicitly,  replacing  $x^M$  by  the  coset space coordinates $x^\m$,
which are some
$D$-invariant combinations of $x^M$  such that
 $$    Z^M_a  H_M^\m = 0 \ , \ \ \  H_M^\m =   {\del x^\m \ov \del x^M } \ ,
\eq{28} $$
then  the metric (25) will  take the form
 $$  G^{\m \n }=  H^\m_M G^{MN}H^\n_N =  H^\m_M {\hat G}^{MN}H^\n_N \ , \ \ \
 $$ $$  G^{\m \n }= {1
\over \kg } ( E^{\m A} E^\n_A   - \g    E^{\m a}E^\n_a ) \ , \  \ \ E^\m_A
\equiv H^\m_M E^M_A
. \eq{29} $$
In the simplest case $H^\m_M = \delta^\m_M$. More generally,
one can choose any set of vectors $H^\m_M$ which are orthogonal to $Z^M_a$.
 As a result, we  will
 get again  eqs. (20)--(22)  with the tensor indices $M,N,...,$  restricted  to
 $G/H$, i.e.
replaced by $\m,\n, ...$.
 Since in the present case (under the constraint)
the  operators $J_A$ are  anti-Hermitian
with respect to  the invariant metric on the coset
$$ \go_{\m\n} = \eta_{ij} E^i_\m E^j_\n \ , \eq{30}  $$
 we find as in (22)  (cf. ref.22)
 $$ \p =  \ha \ln  {\sqrt {G \ov \go } }\ .  \eq{31} $$
 We conclude that the non-triviality
 of the dilaton can be attributed to the fact that  the metric
$G_{\m\n}$ is a `deformed' one, i.e. is different from the standard metric on
the coset.
We have thus  proved that  the combination of the metric and dilaton in (22) is
 equal
to the determinant of $\go$ (and, in particular,  is $k$-independent
$^{13,15,25}$).

It should   certainly be possible also to  compute  the antisymmetric tensor
background  by
comparing the equation for a massless $(1,1)$ state with (12)--(14).   We shall
find the
antisymmetric tensor in  Sec.5  by using  the direct   field-theoretical
approach  where  one
 determines the \sm action  from the gauged WZNW theory (which provides a
Lagrangian
formulation of the coset conformal  theory).    In   Secs.5,6    we  shall
 reproduce  the above expressions for the metric and dilaton  and also,
motivated by the
field-theoretical  derivation, will understand how to  put them  into a more
explicit form.
\vglue 0.6cm
\line{\bf 4.  Effective Action in Gauged WZNW theory  \hfil} \vglue 0.4cm
The classical  gauged  WZNW action
 $$  S= k I(g,A)  \ , \ \ \ I(g,A) = I(g)  +{1\over \pi }
 \int d^2 z \Tr \bigl( - A\,\bd g g\inv +
 \bar A \,g\inv\del g + g\inv A g \bar A  - A \A \bigr)\ , \eq{32} $$
$$  I(g) =   {1\over 2\pi }
\int d^2 z  \Tr (\del g^{-1}
\bd g )  +  {i\over  12 \pi   } \int d^3 z \Tr ( g^{-1} dg)^3   \ ,  \
\eq{33}
$$
is invariant under  the  standard  vector $H$-gauge transformations
(with the generator corresponding to (23))
 $$ g \ra u^{-1} g u \ , \  \ A \ra u\inv ( A + \del ) u \ , \ \
 \A \ra u\inv ( \A + \bd ) u \ , \ \  \ \ u = u (z, \bar z)  \ .$$
The $2d$ gauge  field (with values in the adjoint representation of a subgroup
$H$) at the classical
level plays the role of a Lagrange multiplier, which sets the $H$-components of
the currents to zero.
To obtain a \sm action for gauge-invariant degrees of freedom  out of (32) one
needs to  fix a gauge
and to integrate out the  non-propagating `Lagrange multiplier' degrees of
freedom $A_m$. This   is
straightforward to do  at  the semiclassical level$^6$.
However, integrating out the gauge field $A_m$  while keeping $g$ as  a
background
breaks down conformal invariance starting at two loops$^8$.
To preserve conformal invariance  (which is manifest in the underlying coset
conformal theory)
one must treat $A_m$ and $g$ on an equal footing at the quantum level.

 The idea of how to go beyond the
semiclassical approximation  is to  derive first the quantum effective action
for the theory (32),
solve for the gauge field (and fix a gauge) and identify the result with the
effective action of
the   underlying sigma model$^{14}$.  The  classical \sm action (i.e. its
couplings) are then found by
separating the local (second-derivative) part of the effective action.  Since
the quantum theory of
(32)  can be formulated in  an  exactly conformally invariant way, the
corresponding \sm
couplings  should   also   represent   an exact solution of the \sm conformal
invariance conditions.

As in the case of the partition function$^5$, the  effective action of a gauged
$G/H$  WZNW theory can be represented in terms of the  effective actions of
the  ungauged
 theories for the group $G$ and subgroup $H$. Defining the effective action in
the  WZNW theory by
 $$   \ee{- \G (g) }  =
\int [d\tg] \  \e{ - S(\tg) }  \ \delta [  \J (\tg) - \J  (g) ]
 \ \ ,  \ \ \ \ \J \equiv \bd g g^{-1}  \  , \eq{34} $$
 one finds$^{14,16}$
$$  \G (g) = (k + \half c_G) I(g')  \  , \ \ \
\ \bd g' g'^{-1}=  {k \over k + \ha c_G } \bd g g^{-1} \ , \eq{35} $$
i.e. up to the field renormalisation (which makes $\G (g)$ non-local)
the effective action is given by the classical WZNW action with the shifted
$k$.
This action has the right symmetries (conformal and chiral $G\times G$
invariance) one would like
to preserve at the quantum level. Note that  (35) coincides with $I(g)$ in the
classical limit and
is {\it different} from the usual Legendre transform of the generating
functional $W$ of the
currents  (which contains the unshifted $k$ and is given by the classical
contribution$^{29}$). The
functional $\G$ can be considered as a `quantum' Legendre transform of $W$ in
which almost all of
quantum corrections  except the `one-loop' determinant (providing the shifts of
$k$  in (35))
are absent (cf. ref.29).

  Although it is not obvious   that  the functional (34)  is equivalent to the
 standard
 generating functional of 1-PI  correlators  of the field $g$ itself,  the
resulting action (35)
 is perfectly consistent  with the presence of the shifted $k$
in the quantum    equations of motion    and the  stress  tensor (cf.(15),(17))
in the operator approach  to  WZNW model as   conformal  theory$^{27}$,
$$ (k + \half c_G) \del g (z,\bar z) = \ :J_A(z)  g (z,\bar z)T^A:\ , \ \ \ \
 (k + \half c_G) \bd g (z,\bar z) = \ :{\bar J}_A(\bar z) T^A g (z,\bar z) :\ ,
$$ $$
 T_{zz} =  {1 \over \kg } \eta^{AB}:J_A(z)J_B(z): \  . $$
The action (35) can be considered as a `classical' representation of these
quantum relations with
the normal ordering suppressed (note that the  stress tensor should be given by
the variation of
the effective action over the $2d$ metric and  that the field renormalisation
in (35) is  actually
the renormalisation of the current).
  An action  with the same shift of $k$ as in (35)  (also
originating from a determinant) was  discussed  in ref.28  in connection with
the free-field
representation  of the corresponding conformal theory.

Alternatively, one  may start with  the  assumption
 that the effective action $\G (g) $ in the WZNW theory must satisfy conformal
and chiral $G\times
G$ invariance conditions. Then a  natural  (and probably unique) choice for
such $\G$ is  the
classical action itself, up to renormalisations of $k$ and the  current,
$\G (g) = k' I (g') , \  \bd g' g'^{-1}=  Z   \bd g g^{-1}$.  Correspondence
with the  c.f.t.
approach then   fixes $k'= k + \ha c_G$ (and probably fixes also  $Z={k \over k
+ \ha c_G }$).
  A possibility to  find   an  exact expression for the effective  action of
the WZNW theory
should   not be  surprising, given its solubility in the operator approach.

Maintaining  equivalence  between the local field theory and   operator
conformal theory
results is rather subtle and depends on a choice of a particular regularisation
prescription
(which should correspond to a normal ordering prescription in c.f.t.).
As in the case of  the $3d$ Chern-Simons theory$^{39}$  the  one-loop shift of
$k$
in the effective action may happen in one regularisation and not happen in
another one
(see e.g. ref.40).  The absence of a renormalisation of $k$ in the   standard
Legendre transform
of the generating functional for  correlators of currents (which does not
receive loop
corrections$^{29}$) and its presence in the `quantum' Legendre transform (34)
seems  related to
an observation of ref.41 that  similar `quantum' Legendre transform in
$SL(2,R)$ Chern-Simons
theory  relates  two representations (in terms of  affine and Virasoro
conformal blocks) with
`bare' and  renormalised values of $k$.

Observing that  the  gauged action (32) can be put into the form
 $$ I(g,A) = I (h\inv g \bh ) -  I (h\inv \bh)  \  , \ \ \ \ A = h \del h\inv \
 , \ \  \A
= \bh \bd \bh\inv \ , \eq{36} $$
taking  into account the Jacobian of the change of variables, using (35),
dropping out the non-local
terms introduced by field renormalisations  (which are irrelevant for our
problem  of deriving the
corresponding \sm couplings) and expressing the result back in terms of the
original fields
 $g$ and $A,\A$  we get
the following effective action$^{14}$
$$ \G' (g, A) = (k + \half c_G) I (g, A) + \ha  ( c_G -   c_H)  \O (A)
\ .  \eq{37} $$
Here $\O (A)$  is a non-local  gauge invariant functional of $A $ and $  \A $,
$$ \O (A) \equiv I(h\inv \bh) = \o (A) + \W (-\A) + \1p \int d^2 z \Tr (A\A)
   \ \  , \eq{38} $$
$$ \ \o (A)  = I(h\inv) = - \1p \in
\Tr \{ \half A{\bd\over \del} A  - \third A[{1\over \del}A , {\bd\over \del}A]
+  O(A^4)  \}\
, $$
$$ \W(-\A)  =  I(\bh)
 = -\1p \in  \Tr \{\half \A{\del\over \bd} \A  -  \third \A [{1\over \bd}\A,
{\del\over \bd}\A] +
O(\A^4) \} \  . \eq{39}  $$
It may be possible to solve  the equations for $A,\A$ and eliminate  them  from
the action directly
starting with (37). However, one can simplify the problem by observing that
since we are not
interested in the non-local terms,   it is possible first to truncate the
action $\G'$ by dropping
out  terms that are  of cubic and higher order in $A,\A$.  As a result, we
obtain
the following `truncated' effective action$^{16}$
 $$ \G_{tr} (g, A) = (k + \half c_G) I (g, A) + \ha  ( c_G -   c_H) \O_0 (A) \
, \eq{40} $$  $$
  \O_0 (A) \equiv   \1p \in \Tr ( A\A  - \half
A{\bd\over \del} A - \half \A { \del \over \bd}
\A  )  =  \2p \int  \Tr  F {1 \over \del \bd } F  \ , \ \ \ \  F \equiv \bd A-
\del \A \ .
$$
    The action (40)  is exactly equal to (37) in
 the case  when the subgroup $H$ is abelian.  In contrast to  (37)
 the action (40)  is invariant only under the abelian  gauge transformations,
$A\ra
A + \del \epsilon , \ \A\ra \A + \bd \epsilon $, but it is sufficient for our
purposes  to know
that the full gauge invariance  can be  restored by  re-introducing the higher
order non-local
terms.  The truncated action (40) can be represented also in the form
$$  \G_{tr}(g, A) = (k + \half c_G) \big[ I (g) +  \D I (g, A)
\big ] \ , \  \eq{41} $$ $$   \D I (g, A)\equiv {1\over \pi }
 \int d^2 z \Tr \big[ (  - A\,\bd g g\inv +
 \bar A \,g\inv\del g + g\inv A g \bar A  -  A \A )
$$ $$  +  \ha  b  \ ( AQA + \A Q^{-1} \A - 2 A\A)   \big]  \ , \eq{42} $$
$$    b \equiv  - { (c_G- c_H) \over 2(k+ \half c_G) }  \ ,\ \ \ \ \
 Q\equiv {\bd\over \del} \ , \ \ \ Q^{-1} \equiv {\del \over \bd}  \ .
\eq{43} $$
If we formally set here $\del = \bd$ (i.e. $Q=1$) we  obtain the $d=1$ action
of ref.15
which is the  dimensional reduction of the full action (37) (the  higher-order
commutator terms in
(39) do not contribute in the $d=1$ limit). It is easy to see (on
Lorentz-invariance grounds)
that  the $d=1$ action  is,  in principle,   sufficient in order to  extract
 the metric and dilaton  couplings  of the corresponding   sigma model$^{15}$.
However, to
 derive the antisymmetric tensor coupling one should  use the  direct  $2d$
approach
based on (41).

  To   obtain the \sm action from (41)   one should first
solve for  the gauge field  and   then    discard  all the
non-local terms.  This will be discussed  in Sec.5. Note that   it is
not correct  just to omit  the terms with the operator $Q$ insertions  (since
$Q$
has  dimension zero and
since this  would break  the  gauge invariance of (42)); it is  also not
correct     to replace
$Q$ by 1  (this would break the Lorentz invariance).

The above results for  the effective actions have  generalisations$^{16}$ to
the case of $n=1$
supersymmetric (gauged) WZNW theory.
Computation of the effective action in  the  supersymmetric WZNW theory
can be reduced to that  in the bosonic WZNW theory by using the
observation$^{30}$
that  by  a  formal field redefinition the  supersymmetric action  can be
represented   as a  sum  of the bosonic  action and the
action of the  free Majorana fermions in the adjoint representation of the
group $G$.
 The transformation of fermions  which is
needed to decouple
 them from $g$ is, however,  chiral   and  therefore   produces a non-trivial
Jacobian. The
logarithm of the fermionic determinant
gives a contribution  proportional to the  bosonic  WZNW action,  leading  to
  the
shift of the coefficient $k$ in the bosonic part of the action$^{31,32}$:
$ k \ra  {\hat
k}\equiv k- \ha c_G  .$\footnote{*}{\ninerm\baselineskip=11pt
This  shift of $k$ is consistent with  \sm perturbation theory$^{33}$. \hfil}
    As a result, the  effective action in the
ungauged supersymmetric WZNW theory
is obtained by replacing  $k$   by  $ k- \ha c_G$  in (35)$^{16}$
$$\G (g) = k  I(g')  \  , \ \ \ \        \bd g' g'^{-1}=  (1-   { c_G \ov 2k })
\bd g g^{-1} ,
\eq{44} $$
i.e.  is equal to the classical WZNW action with {\it unshifted}  $k$:
the  shift of $k$   in $\G$  produced by integrating  out fermions
     is exactly cancelled
 out by the  bosonic contribution.

In the  case of the gauged supersymmetric WZNW theory  one may  use a
superfield formalism
to maintain  manifest  supersymmetry.\footnote{**}{\ninerm\baselineskip=11pt
This  possibility was  first  mentioned  in ref.32 where,
however, the component approach (in the particular case of $G=H$) was
discussed. \hfil}
Then the path integral quantisation of the
supersymmetric theory  becomes  parallel to that of the bosonic  theory (see
(36) etc.)  with fields
replaced by superfields.  The theory is reduced to that of the  two ungauged
supersymmetric WZNW
theories for $G$ and $H$ (with the only  difference  with respect to  the
bosonic case that now
the Jacobian of the change of variables is trivial).  Applying (44) one finds
the following
expression for the
  (bosonic part of  the)  effective action$^{16}$
 $$   \G_{susy} (g,A) = k I (\tg' ) - k   I (\tth')         \ ,   $$ $$
\ \ \ \   \bd \tg' \tg'^{-1}=  (1-   { c_G \ov 2k })  \bd \tg \tg^{-1} \ , \ \
  \bd \tth' \tth'^{-1}=  (1-   { c_H \ov 2k })  \bd \tth \tth^{-1}\ \   \eq{45}
 $$
(here $\tg=  h\inv g \bh   , \ \tth= h\inv \bh
$; cf.(36)).
As in the ungauged \SWZNW  but in  contrast  with the
bosonic  gauged WZNW
case
 there are  no shifts in the overall coefficients  of the $G$- and $H$-terms
in
$\G_{susy}$.  Ignoring the non-local corrections
introduced by the field renormalisations  we
conclude   that  the local part
of the effective action  of the gauged \SWZNW is equal to the {\it classical}
action of the
bosonic  gauged WZNW theory
  $$  \G_{susy}' (g, A) = k I(g,A) = k \big[ I(g)  +{1\over \pi }
 \int d^2 z \Tr \bigl( - A\,\bd g g\inv +
 \bar A \,g\inv\del g    $$ $$ +   g\inv A g \bar A  - A \A \bigr) \big] \   ,
\eq{46}
$$
i.e. in contrast  with  the  bosonic case (37),  it does not contain the
quantum correction  term proportional to $ b = - { (c_G - c_H) \over 2(k+ \half
c_G) }$.
As a consequence,  the exact form of  the  corresponding \sm will be
equivalent to the
`semiclassical' form of the \sm corresponding to the bosonic theory (with no
shift of $k$).
 This conclusion is the same as the one
 reached   in the operator approach in ref.10 (in the case of the
$SL(2,R)/U(1)$
supersymmetric  model) and in ref.13 (in the case of a  general $G/H$
supersymmetric theory).

 \vglue 0.6cm
\line{\bf 5.  Derivation of Sigma Model Couplings  \hfil}
 \vglue 0.2cm
\line{\bf
 \  \ \ \ in Field-Theoretical Approach \hfil} \vglue 0.4cm
As explained above, our starting point  will be  the truncated effective action
(41),(42).
Since it is quadratic in the gauge potentials $A,\A$  it is  straightforward to
integrate them out.  Representing (42)  as
 $$  \D I (g, A)= {1\over \pi }
 \int d^2 z \  \big[ (  - A\J  +
 \bar A J)   +    A  N \A
 +  \ha  b  \ ( AQA + \A Q^{-1} \A )   \big]  \ , \eq{47} $$
 $$   J_a =  \Tr (T_a g\inv\del g) \ , \   \ \J_a = \Tr (T_a \bd g g\inv ) \ ,
\ \ \ \ \  N_{ab}
\equiv M_{ab}  - b  \eta_{ab}\ , $$ $$  \  \ M_{ab} \equiv C_{ab} - \eta_{ab} \
, \ \ \ \
  C_{ab } \equiv \Tr (T_a g T_b g\inv ) \ , \ \ \ \  \Tr (T_aT_b) = \eta_{ab}
\ , \eq{48} $$
and  solving for  $A,\A$   we obtain (after
omitting the   non-local terms in which  $Q$ or $Q^{-1}$ are  acting on $N$)
$$   \D I (g) ={1\over 2 \pi }
 \int d^2 z \ \big[ J \V^{-1 } (N^T \J + b Q J)+ \J V^{-1 T} (N J + b
Q^{-1}\J)\big]   \ ,
 \eq{49} $$
$$ V\equiv N N^T - b^2  = MM^T- 2bM_S\ , \  \eq{50} $$ $$  \ \ \V\equiv N^T N -
b^2= M^TM
-2bM_S \ ,\ \ \ \ M_S\equiv \ha (M + M^T) \ . $$
Ignoring   the non-local  terms,  we can replace  $ Q^{-1}\J$ by $ \Tr (T_a
\del
g g\inv ) $
and  $ QJ$ by $ \Tr (T_a g\inv\bd g ) $.     Using the  parametrisation in
terms of the group-space coordinates $x^M$ (we shall fix the gauge by
restricting $x^M$ to
coset space coordinates $x^\m$ later)
$$ g\inv \del g = T_A E^A_M (x)  \del x^M \ , \ \ \ g\inv \bd g = T_A
E^A_M(x)
 \bd x^M \ ,  \ \ \  \del g  g\inv  = T_A \E^A_M (x)  \del x^M \ , $$ $$    \bd
g g\inv = T_A
\E^A_M (x) \bd x^M \ ,\ \ \  \  \E^A_M =  C^A_{ \ B }(x) E^B_M \ , \ \ \ \
C_{AB} = \Tr ( T_A g T_B
g\inv ) \ , \eq{51} $$
 we  can put  the local  part of
the action (41),(49) into the \sm form$^{16}$
 $$
\G_{loc}(g)=  - {1 \over \pi \a' } \int d^2 z \ {\cal G}_{MN} (x) \del
x^M \bd x^N \ , \ \ \ \
\a' = {2\ov k + \ha c_G} \ , \eq{52} $$
$$ G_{MN}   \equiv  {\cal G}_{(MN )}= { G}_{0MN }  -
 b  (\V^{-1 })_{ab} E^a_M E^b_N -
  b (V^{-1 })_{ab} \E^a_M \E^b_N  $$ $$ -
 2 (  \V^{-1 } N^T )_{ab} E^a_{(M} \E^b_{N)}\ , \eq{53} $$
$$  B_{MN}   \equiv  {\cal G}_{[MN ]} =B_{0MN} -
 2 (  \V^{-1 } N^T )_{ab} E^a_{[M} \E^b_{N]}\ . \eq{54} $$
  Here ${ G}_{0MN }$ , ${ B}_{0MN }$   stand for  the  original  WZNW
couplings,
$$ G_{0MN} =\eta_{AB} E^A_M  E^B_N = \eta_{AB} \E^A_M  \E^B_N \ , \ \ \  $$ $$
 3\del_{[K} B_{0MN]} =  E^A_KE^B_ME^C_N f_{ABC} =\E^A_K\E^B_M\E^C_N f_{ABC}
\ . \eq{55}  $$
As in refs.6,14,  the  local part of the determinant of the  matrix in the
quadratic  ($A,\A$) term in (47) gives the  dilaton  coupling
$$  \p   = \p_0  - \fourth  {\ \rm ln \ det \ }  V   \ . \eq{56 } $$
The  same expressions  for the metric (53) and dilaton (56) were  found   using
the $1d$
 reduction  of the action (37) in ref.15.   The  result for the antisymmetric
tensor
coupling (54) is equivalent to the expression   also  suggested   in  ref.15
on the basis of
an analogy with the  expression  for the
metric.\footnote{*}{\ninerm\baselineskip=11pt
  Ref.15 contains also a derivation
(without assuming the $1d$ reduction) of the antisymmetric tensor coupling in a
  particular case of
the  $SL(2,R)\times SO(1,1)/SO(1,1) $  ($D=3$ `black string')  model.\hfil}

The   expressions for the metric  and the antisymmetric tensor  (53),(54)  are
yet
in rather abstract  form.  To   give a more explicit and useful representation
for the metric one
should first  express $\E^a_M$  in terms of $E^a_M$ and $E^i_M$ with the help
of (51)
$$ \E_{aM} = C_{ab} E^b_M + C_{ai} E^i_M \ ,  \ \  \ C_{ad} C_{b}^{\ d } +
C_{ai} C_{b}^{\ i}  =
\eta_{ab} \ , \ \ \  E^A_M E^M_B = \delta^A_B \ ,  \ \ E^A_M E^N_A = \delta^N_M
\  \eq{57}  $$
(the  indices are raised and lowered with the help of  $\eta_{AB}$ and
$G_{0MN}=
E^A_ME_{AN}$).  Then (53)  reduces to
$$ G_{MN}  =  h_{AB} E^A_M E^B_N   =
  h_{ij} E^i_M E^j_N +  ... \ , \ \ \ \
 h_{ij }= \eta_{ij}   -   b  V^{-1 }_{ab}  C^a_{\ i} C^b_{\ j}  \ . \eq{58} $$
It is  now straightforward  to check that the metric (53),(58)  is degenerate,
having $D_H$ null vectors
$$    Z^N_a =  E^N_a - \E^N_a = - M_{ab} E^{Nb}- C_{ai} E^{Ni} \ , \ \ \ \
G_{MN} Z^N_a =0 \ .
\eq{59} $$
 These vectors  are recognised as being the generators of the vector
subgroup $H$ of  the $G\times G$ symmetry of the WZNW  action  which was gauged
in (32) (cf.(23)).

 To obtain  a non-degenerate metric  one  should restrict  $G_{MN}$  to the
subspace
orthogonal (with respect to $G_{0MN}$)
to  the null vectors $Z^N_a$.
 One  can  change  the original basis $E^M_A = (E^M_i , E^M_a)$  to  a  new
one
$ (H^M_i, Z^M_a)$  with $H^M_i$ being  orthogonal to $Z^M_a$.
Then  the  degenerate metric (58)  takes the form ($H^i_M \equiv H^N_j
\eta^{ij} G_{0MN}$)
$$ G_{MN}  =  g_{ij} H^i_M  H^j_N \ , \ \ \ \  \  G_{0MN} H^M_i Z^N_a= 0 \ , \
 \ \Pi^N_M H^M_i =
H^N_i \ , \eq{60} $$
where the projection operator $\Pi$ is the same as in (24) and  the  expression
for $g_{ij}$ depends
on a particular choice of the vectors $H^i_M$
 (there is a freedom of making a transformation $H^M_i \ra  \Lambda_i^j
{H}^M_j$).
A  simple  choice of  $H^M_i$ is
(we shall use bars to denote objects  corresponding  to this basis)
$$  \bH^i_M = E^i_M - M^{-1}_{ab} C^{bi} E^a_M = p^i_j E^j_M +
M^{-1}_{ab} C^{bi} M^{-1a}_{\ \ c} Z^c_M\ , \eq{61} $$ $$ \  \ \  p_{ij} \equiv
\eta_{ij} +
(MM^T)^{-1}_{ab}  C^a_{\ i} C^b_{\ j}\  \ .
 $$
Then
 $$G_{MN}    =
  {\bar g}_{ij} \bH^i_M \bH^j_N\ , \ \ \ \
{\bar g}_{ij} = h_{ij} = \eta_{ij} -   b  V^{-1 }_{ab} C^a_{\ i} C^b_{\ j}\ .
 \eq{62} $$
Since the inverse and determinant of the  $D\times D$ matrices  of generic
 form  $m_{ij }= \eta_{ij} +  f_{ab}
C^a_{\ i} C^b_{\ j}$  are given by
$$ m^{-1}_{ij} = \eta_{ij} + f^{(-1)}_{ab}  C^a_{\ i} C^b_{\
j}\ , \ \  \
f^{(-1)}=- [f^{-1}  +  ( 1- CC^T) ]^{-1}\ , $$ $$
\ \ \  \det m =  \det [1+ f(1-CC^T)]  \ , \eq{63} $$
 we find (cf.(25)--(27))
$$  G^{-1MN}=  \Pi^M_K {\hat G}^{-1KL} \Pi_L^N \ ,  \ \ \  G^{-1MN}G_{NK} =
\Pi^M_K \ ,  $$
$$ {\hat G}^{-1KL}=   E^M_AE^{AN} -  \g   E^M_aE^{aN}
= E^M_iE^{iN}  -   (\g -1)   E^M_aE^{aN}\ , \eq{64} $$ $$ \ \ \
 \g = (b+1)^{-1} = {k+\ha c_G\over  k+\ha c_H } \ ,
  $$
$$\det {\bar g}_{ij} = \det [ (1+b) V^{-1}] \ \det  M^2 \ .  \eq{65} $$
The inverse of the metric (62) thus has a  much simpler structure
than  the metric itself.

Since the \sm  on the  whole  group space $ \int d^2z \  G_{MN} (x) \del  x^M
\bd x^N  + ... $ has
the  gauge invariance (generated by $Z^M_a$) the   final step  is to  fix a
gauge, e.g.
restricting  coordinates  $x^M$ on $G$ to coordinates  $x^\m$ on $G/H$.  Let  $
R^a (x^M) =0 \ , \ \
R^a_M \delta x^M  = 0 \ , \ \    R^a_M \equiv  \del_M R^a  $
 be  a gauge condition (the corresponding ghost determinant  that should be
included in the
measure is  $\det R^a_M Z^M_b $).  One  may  either  add a gauge term into the
\sm action
(which will then  depend on  all $x^M$ coordinates) or solve explicitly the
gauge condition,
 expressing  $x^M$
  in terms of $D$ coordinates $x^\m$ on $G/H $, $\ x^M = x^M (x^\m ).$
\footnote{*}{\ninerm\baselineskip=11pt {If one uses  the formulation in terms
of all  $D_G$
coordinates $x^M$
one should  also impose as usual the gauge invariance    (BRST invariance)
condition  on the
observables. Adding a gauge-fixing term in the action one  obtains the
following  non-degenerate
metric on  $G$: $\  {\bar
 G}_{MN} = G_{MN} + q_{ab} R^a_M R^b_N$.
 The  determinant of the degenerate metric
  $G_{MN} $ is defined as   $ ({\rm det\ } { G_{MN}})^{-1/2} =  ({\rm det\ }
 {{\bar G}_{MN}})^{-1/2} \ {\rm det\ }  (R^a_M Z^M_b)  (\det q_{ab})^{1/2}\
. $ }}
In the latter case we  will get a \sm on the $D$-dimensional space  with
$E^A_M$   replaced by  the $D\times D $
matrix   $E^A_\m \equiv E^A_M \del_\m x^M $, i.e.
 $H^i_M$   replaced by  the $D\times D $
matrix   $H^i_\m $ (i.e.  a   vielbein),  $G_{MN}$  replaced by $G_{\m\n}$,
etc.
 The  expressions for the  \sm metric  and antisymmetric tensor  then are
$$ G_{\m\n}    =
  {\bar g}_{ij} \bH^i_\m \bH^j_\n \ , \ \ \ \ \bH^i_\m = \bH^i_M \del_\m x^M \
,\eq{66} $$
$$  B_{\m\n}  = B_{0\m\n} -
 2 (\V^{-1 } N^T  C)_{ab} E^a_{[\m} E^b_{\n]}
-2 (\V^{-1 } N^T)_{ab}C^b_{\ i}  E^a_{[\m} E^i_{\n]}
\ . \eq{67} $$

\vglue 0.6cm
\line{\bf 6. Discussion  \hfil} \vglue 0.2cm

Let us now compare the   above expressions  for the background metric  (66) and
 dilaton (56)
corresponding  to  the gauged WZNW model
 with the
results  (29),(31) which were  found  in Sec.3  by identifying the operator
$L_0$
of the $G/H$   coset conformal theory with a
`Klein-Gordon' operator in a background.
Using (64) we  conclude that the inverse of the metric (62),(66) is equivalent
to the inverse
  metric (25),(29) found in the operator approach (up to the overall factor
$\kg$ which we
absorbed in $\a'$ in (52)). We have thus  proved  that  both the operator  and
the
field-theoretical approaches  lead to the same expression for the target space
metric.

Though  expected, this  equivalence  is realised in  a rather non-trivial way.
 While in the  operator approach  one obtains  naturally the  {\it inverse} of
the metric,
the field-theoretical  derivation   or sigma model approach   gives   the
metric itself.
 The
two approaches  are in a sense `dual' like    momentum and coordinate
representations
 or Hamiltonian and Lagrangian formalisms.
The procedure of elimination of the $H$-gauge field  is  analogous
to that of integration over the momentum (this analogy can  probably  be made
more precise by
rewriting the WZNW term in the effective action (37) in terms of  an  auxiliary
`momentum'
or $G$-gauge field variable and extracting the {\it inverse} of the metric from
the  term
quadratic in
all components of the gauge field).

The metric (6.1) can be considered as  a
 `deformation' of the `round' metric on $G/H$.
The latter  corresponds to the \sm which is found by integrating out the gauge
field
with values in the algebra of $H$ in the  action
 invariant under the $H$-gauge transformations  generated by $E^a_M$, i.e.
$g'=gu$
 $$ I = \int d^2 z   \Tr ( g\inv \del_m g +  A_m )^2  \ . \eq{68} $$
This action (and hence the resulting \sm metric) has also global $G$-invariance
which  is {\it absent} in the gauged WZNW action (46) (being broken by the
$g^{-1}Ag\A$-term).
Before gauge fixing,  we
get a degenerate metric   $G^{(0)}_{MN}$ on the  full $G$
 (with null vectors
$E^M_a$). Solving  a gauge condition  $R^a(x^M)=0$
 and expressing $x^M$ in terms of $x^\m$ we obtain
the  metric $G^{(0)}_{\m\n}$ on  the $D$-dimensional coset  space $G/H$,
$$ G^{(0)}_{MN} =\eta_{ij}E^i_M E^j_N \ , \ \ \ \  G^{(0)}_{\m\n}=
\eta_{ij}E^i_\m E^j_\n  \ , \ \ \ E^i_\m = E^i_M \del_\m x^M
 \ . \eq{69} $$
Noting  that   according to (61) $ \bH^i_M = E^i_M + O(E^a_M)$  and    choosing
the gauge condition
  such that $ \del_M R^a = E^a_M $
 one can show  that
$$ \det G_{\m\n} =   \det G^{(0)}_{\m\n} \ \det {\bar g}_{ij} \ ({\rm det}\
M)^{-2}\ ,  \eq{70} $$
$$ \det G^{(s)}_{\m\n}  = \det G^{(0)}_{\m\n} \  ({\rm det}\ M)^{-2}  \ ,  \ \
\ \
 G^{(s)}_{\m\n}= {\eta}_{ij} \bH^i_\m \bH^j_\n \ ,\eq{71} $$
where $ G^{(s)}_{\m\n}$ is the semiclassical ($b=0$) limit of $ G_{\m\n}$ and
$\det M $  is the
corresponding `ghost determinant' ($E^a_M Z^M_b = - M^a_b ,$  see (59)).

Using (70),(71),(65)  and the expression for the dilaton (56) one finds
$$ \sqrt { \det G_{\m\n} }  \ \e{-2 \p} =
 \sqrt {  \det G^{(s)}_{\m\n} } \ \sqrt { \det {\bar g}_{ij} }
 \ \e{-2 \p}
 $$ $$ = \  a \  \sqrt {
 \det G^{(s)}_{\m\n} } \ \det M  =\  a  \ \sqrt {  \det G^{(0)}_{\m\n} } \  ,
\eq{72} $$
$$ a =  (\sqrt {1+b} )^{D_H}\e{-2 \p_0} \ . $$
The constant $a$ can be made equal to $1$ by a choice of $\p_0$.
Therefore, in agreement with the  result of the operator approach (31),
the dilaton is given   by the
logarithm of the ratio  of the determinants of the  metric and the  invariant
metric on the coset space, i.e.
  the `measure factor' (72) is  nothing but the
canonical  measure on   $G/H$.\footnote{*}{\ninerm\baselineskip=11pt
The   fact    that  the product $\sqrt G \ {\rm
e}^{-2 \p}$  is  $k$-independent   was  first  observed  in the $SL(2,R)/U(1)$
case$^{9,25}$.
 It  was   further checked$^{13}$  on a number
of  non-trivial  $G/H$ models. Refs.13,15   formulated  this fact  as a general
statement
  and gave arguments supporting it  using path integral measure considerations.
\hfil }

Corresponding to the coset $G/H$ conformal theory, the  \sm with the metric
(66), antisymmetric
tensor (67) and dilaton (56) should  be   conformally
invariant to all orders in the  loop expansion, i.e. should represent a large
class of exact
solutions of string equations.  Depending on $b$, the  fields $G_{\m\n},
B_{\m\n}$
and $\p$ are
non-trivial functions of  the parameter $k$ or  $\a'$ (see (43),(52)),
$  k +\ha c_G = {2\ov \a'}    \ ,   \ \ b=   {1\over 4}(c_H - c_G) \a' ,
$
 (the semiclassical limit  is $b \ra 0)$.
 The dependence of the metric  (66),(62)  on $\a'$  can be represented in the
following   symbolic way:
 $ G = G^{(s)}  + {\a' F_1 \over F_2 + \a' F_3 }$. At the same time,  the
inverse metric  (64),(26)
 which
appears naturally in the operator approach    has simpler dependence:
 $ G^{-1} = (G^{(s)})^{-1}   + {\a'  \over  \a' +  q} F_4\ , \  q=
4/(c_H-c_G)$.

In spite of the  fact that  the underlying conformal theory has $G/H$
structure,  the corresponding
symmetry   is not explicit at the \sm level.  It should be emphasised that it
is the condition of
conformal invariance that makes the resulting metric a `deformed'
(non-symmetric)
one.\footnote{*}{\ninerm\baselineskip=11pt  Similarly,  in the operator
approach  even though one
uses the affine $G$-symmetric framework it is the condition of conformal
(Virasoro) invariance (16)
that, in general, constrains $\C^{AB}$  in such a way that the corresponding
target space metric is
non-symmetric.\hfil }
  In fact, the  standard $G/H$ \sm
(68)  is not conformal since the metric $\go$ of the homogeneous space  has a
non-vanishing Ricci
tensor, giving a  non-vanishing  one-loop  $\b$-function.  One possibility to
make a conformal \sm
out of the homogeneous space metric  is  to introduce two extra dimensions (one
space-like and one
time-like) and impose the null Killing symmetry condition on the resulting
$(2+D)-$dimensional metric
 as discussed in Sec.2.\footnote{**}{\ninerm\baselineskip=11pt   One can also
add just one extra
dimension (time) and compensate the conformal anomaly by the time evolution of
the scale
factor (and dilaton). Such `cosmological' solutions  (with $G/H$ being a
sphere) were found
to exist  in  the order $\a'$ approximation$^{34}$ but it is not clear how to
extend them to exact solutions.\hfil }
  Another  possibility -- which is realised in the models  corresponding to the
coset conformal theories -- is to deform the metric, introducing at the same
time a non-trivial
dilaton (and antisymmetric tensor)
background.\footnote{***}{\ninerm\baselineskip=11pt
It is an interesting question if there are other exact solutions  of the \sm
conformal invariance
conditions  for  which the  homogeneous space metric remains  undeformed but
the conformal anomaly
is cancelled out by the contribution of the antisymmetric tensor (cf. ref.35).
\hfil }

As we have found in Sec.4,  the  local part  of  the bosonic term in  the
effective action of  the
gauged $n=1$ supersymmetric WZNW theory  is  equal to the classical bosonic
gauged WZNW
action (46) with
unshifted $k$ and no `quantum' $b$-term. Thus  in the supersymmetric case
$\a' = {2/k}$  and
the corresponding  exact \sm couplings are given   by  the `semiclassical'
limit $b=0$ of the
bosonic WZNW theory   expressions
(66),(67),(56)
$$ G^{(s)}_{\m\n}    =
  {\eta}_{ij} \bH^i_\m \bH^j_\n = {\eta}_{ij}( E^i_\m - M^{-1}_{ab} C^{bi}
E^a_\m)
( E^j_\n - M^{-1}_{ab} C^{bj} E^a_\n) \ , \ \ \ \eq{73} $$ $$
  B^{(s)}_{\m\n}  = B_{0\m\n} -
 2 (M^{-1})_{ab} E^a_{[\m} E^b_{\n]}
-2 (M^{-1})_{ab}C^b_{\ i}  E^a_{[\m} E^i_{\n]}
 \  ,  \  \ \  \  \p^{(s)} =\p_0 - \ha \det M \
 . $$
The non-triviality of $ G^{(s)}_{\m\n} $ is `hidden' in the choice of
$\bH^i_\m$ (see (61)).
  For example, in the case when $G/H$ is K\"ahler, eq.(73)
gives the couplings of
the \sm  corresponding to a class of $n=2$   superconformal theories$^{36,13}$.
 The  fact that the \sm couplings in the supersymmetric case  do not depend on
$\a'$  means  that they  solve the  \sm conformal invariance conditions
 at each order of the loop expansion. Namely, once the one-loop  conditions are
satisfied,   all
higher-loop corrections to the  $\beta$-functions  should
 vanish  (up to a field redefinition ambiguity) on
the corresponding   background.\footnote{****}{\ninerm\baselineskip=11pt
This was checked explicitly up to the 5-loop order
in the supersymmetric $SL(2,R)/U(1)$ model$^{10}$. While the  two -  and three
- loop   terms in the
$\b$-function of the $n=1$ supersymmetric \sm  are known to  vanish$^{37}$  (in
the minimal subtraction
scheme), the four-loop term does not vanish  in general$^{38}$.  However, there
exists such a
renormalisation scheme in which it vanishes for the  `one-loop' $D=2$
background of refs.7,6.
\hfil }  It is interesting to note  that these models  have, in general, only
$n=1$ (and not $n=2$
or $n=4$)  supersymmetry (as we have discussed in Sec.2, the models of ref.3
represent another
example of finite $n=1$ supersymmetric models).

Though particular examples of geometries corresponding to coset conformal
theories
look  complicated and unusual (see ref.13 and refs. there), they may actually
be more characteristic to string
theory than more symmetric  spaces (which are not string solutions).
Having  behind them a  deep algebraic  structure   of
  coset conformal theory,
they may have some interesting universal features.
The representation of the background fields in terms of the group vielbeins
(66),(67)  generalises similar expressions for the group spaces
and may serve as a basis for a study  of  their geometrical properties.

\vglue 0.6cm
\line{\bf 6. Acknowledgements \hfil}
\vglue 0.4cm
I am grateful to  L. Alvarez-Gaum\'e, E. Kiritsis and C. Kounnas  for
discussions and
interesting remarks.
   This work was partially supported by a grant from SERC.
\vglue 0.6cm
\line{\bf 7. References \hfil}
\vglue 0.4cm
\medskip
\item {1.}  C. Lovelace, \pl B135(1984)75; \np B273(1986)413;
C. Callan, D. Friedan, E. Martinec and M. Perry, Nucl.Phys. B262(1985)593.
\item {2.} A.A. Tseytlin, \pl B288(1992)279; \np B390(1993)177.
\item {3.} A.A. Tseytlin, preprint Imperial/TP/92-93/7 (1992), Phys. Rev. D47
(1993) no.8.
\item {4.} K. Bardakci and M.B. Halpern, \pr D3(1971)2493;
M.B. Halpern, \pr D4(1971)2398;    P. Goddard,
A. Kent and D. Olive, \pl B152(1985)88; \cmp 103(1986)303.
\item {5.}  D. Karabali, Q-Han Park, H.J. Schnitzer and
Z. Yang, Phys. Lett. B216 (1989)
 307;  D. Karabali and
 H.J. Schnitzer, \np B329 (1990) 649;
K. Bardakci, E. Rabinovici and
B. S\"aring, \np B299(1988)157;
 K. Gawedzki and A. Kupiainen, \pl B215(1988)119;
\np B320(1989)625.
\item {6.}   E. Witten, \pr D44(1991)314;  K. Bardakci, M. Crescimanno and E.
Rabinovici, \np
B344(1990)344.
\item {7.} S. Elitzur, A. Forge and E. Rabinovici, \np B359(1991)581.
\item {8.} A.A. Tseytlin, \pl B264(1991)311.
\item {9.} R. Dijkgraaf, H. Verlinde and E. Verlinde, \np B371(1992)269.
\item {10.} I. Jack, D.R.T.  Jones and J. Panvel, Liverpool preprint
LTH-277 (1992), Nucl.Phys. (1993).
\item {11.}  M.J.  Perry and E. Teo, preprint DAMTP R93/1 (1993); P. Yi,
preprint CALT-68-1852
(1993).
\item {12.}  A.A. Tseytlin and C. Vafa, \np B372(1992)443.
\item {13.}  I. Bars and  K. Sfetsos, \pr D46(1992)4510; preprint USC-92/HEP-B3
(1992);
 K. Sfetsos,  preprint USC-92/HEP-S1 (1992).
\item {14.}  A.A. Tseytlin, preprint Imperial/TP/92-93/10 (1992), \np (1993).
\item {15.}  I. Bars and  K. Sfetsos,  preprint USC-93/HEP-B1 (1993).
\item {16.} A.A. Tseytlin, preprint CERN-TH.6804/93 (1993).
\item {17.}  S.R. Das, S. Naik and S.R. Wadia, \mpl
A4(1989)1033;
J. Polchinski, \np B324(1989)123;
 T. Banks and J. Lykken, Nucl.
 Phys. B331(1990)173;  A.A. Tseytlin, \ijmp A5(1990)1833.
\item {18.} V. Novikov, M. Shifman, A. Vainshtein and V. Zakharov, \pl
B139(1984)389; Phys. Rep. 116(1984)103;
A. Morozov, A. Perelomov and M. Shifman, \np B248(1984)279;
M.C. Prati and A.M. Perelomov, \np  B258(1985)647.
\item {19.} B. Zumino, \pl B87(1979)203;
L. Alvarez-Gaum\'e and   D. Freedman, \pr D22(1980)846.
\item {20.} P. Di Vecchia and  S. Ferrara, \np B130(1977)93;
 E. Witten, \pr D16(1977)2991.
\item {21.} M.B. Halpern and E.B. Kiritsis, Mod. Phys. Lett.

 A4(1989)1373;  A4(1989) 1797 (E).
\item {22.} A.Yu. Morozov, A.M. Perelomov, A.A. Rosly, M.A. Shifman and A.V.
Turbiner, \ijmp
A5(1990)803.
\item {23.} M.B. Halpern, E.B. Kiritsis, N.A. Obers, M. Porrati and J.P.
Yamron,
\ijmp A5(1990)2275;
  A.Yu. Morozov,  M.A. Shifman and A.V. Turbiner, \ijmp
A5(1990)2953;
A. Giveon, M.B. Halpern, E.B. Kiritsis and  N.A. Obers,
\np B357(1991)655.
\item {24.} M.B. Halpern and   J.P. Yamron, Nucl. Phys. B332(1990)411;
Nucl. Phys.
B351(1991)333.
\item {25.} E. Kiritsis, \mpl A6(1991)2871.
\item {26.} E. Witten, \cmp 92(1984)455.
\item {27.} V. Knizhnik and A. Zamolodchikov, \np B247(1984)83.
\item {28.} A. Gerasimov, A. Morozov, M. Olshanetsky, A. Marshakov and
S. Shatashvili, \ijmp
A5(1990)2495.
\item {29.} H. Leutwyler and M.A. Shifman, \ijmp A7(1992)795.
\item {30.} P. Di Vecchia,  V. Knizhnik, J. Peterson and P. Rossi,
 \np
B253(1985)701; R. Rohm,  \pr D32(1985)2845;
 H.W. Braden,
\pr D33(1986)2411.
\item {31.} A.N. Redlich and H.J. Schnitzer, \pl B167(1986)315;
 B193(1987)536 (E);
E. Bergshoeff, S. Randjbar-Daemi, A. Salam,
 H. Sarmadi and E. Sezgin,
\np B269(1986)77;  A. Ceresole,
 A. Lerda, P. Pizzochecco
 and
P. van Nieuwenhuizen, \pl B189(1987)34; J. Fuchs, \np B286(1987)455;

\np B318(1989)631.
\item {32.}  H. Schnitzer, \np B324(1989)412.
\item {33.} R.W. Allen, I. Jack and D.R.T. Jones, Z. Phys. C41(1988)323.
\item {34.} A.A. Tseytlin, \ijmp D1(1992)223.
\item {35.} L. Castellani and D. L\"ust, \np B296(1988)143.
\item {36.} Y. Kazama and H. Suzuki, \np B321(1989)232; \pl B216(1989)112.
\item {37.} L. Alvarez-Gaum\'e,   D. Freedman and S. Mukhi, \ap 134(1981)85;
L. Alvarez-Gaum\'e, \np B184(1981)180.
\item {38.} M.T. Grisaru, A. van de Ven and D. Zanon, \np B277(1986)409.
\item {39.}  E. Witten, \cmp 121(1989)351; G. Moore and N. Seiberg,
\pl
B220(1989)422; L. Alvarez-Gaume, J. Labastida and A. Ramallo, \np
B334(1990)103.
\item {40.} M.A. Shifman, \np B352(1991)87.
\item {41.} H. Verlinde, \np B337(1990)652.
\end

\end